\documentclass[10pt]{article}
\usepackage{graphicx}
\usepackage{amsmath,amssymb}
\newcommand{\rmd}{\mathrm{d}}
\newcommand{\rmi}{\mathrm{i}}
\newcommand{\bi}[1]{\textbf{\textit{#1}}}
\begin{document}
\title{\textbf{Linear, diatomic crystal: single-electron states and large-radius excitons}}
\author{Vadym M. Adamyan\footnote{E-mail: vadamyan@onu.edu.ua}, Oleksii A. Smyrnov\footnote{E-mail: smyrnov@onu.edu.ua}\\
\emph{Department of Theoretical Physics, Odessa I. I. Mechnikov
National University,}\\ \emph{2 Dvoryanskaya St., Odessa 65026,
Ukraine}}\date{January 2, 2009}\maketitle
\begin{abstract}
The large-radius exciton spectrum in a linear crystal with two
atoms in the unit cell was obtained using the single-electron
eigenfunctions and the band structure, which were found by the
zero-range potentials (ZRPs) method. The ground-state exciton
binding energies for the linear crystal in vacuum appeared to be
larger than the corresponding energy gaps for any set of the
crystal parameters.
\end{abstract}
PACS number(s): 73.22.Dj, 73.22.Lp, 71.35.Cc

\section{Introduction}
\setcounter{equation}{0}

The study of the quasione-dimensional semiconductors with the
cylindrical symmetry became an urgent problem as soon as
investigations of semiconducting nanotubes had been launched. One
of the most important trends of research in this field is the
study of optical spectra of such systems, which should include the
exciton contributions~\cite{fluor}-\cite{ppv}. Evidently, the
quasione-dimensional large-radius exciton problem can be reduced
to the 1D system of two quasi-particles with the potential having
Coulomb attraction tail. Due to the parity of the interaction
potential the exciton states should split into the odd and even
series. In~\cite{adsmyr} we show that for the bare and screened
Coulomb interaction potentials the binding energy of even excitons
in the ground state well exceeds the energy gap (in the same work
we also discuss the factors, which prevent the collapse of
single-electron states in isolated semiconducting single-walled
carbon nanotubes (SWCNTs). But the electron-hole (e-h) interaction
potential and so the corresponding exciton binding energies may
noticeably depend on the electron and hole charge distributions.
So it is worth to ascertain whether the effect of seeming
instability of single-electron states near the gap is inherent to
the all quasione-dimensional semiconductors in vacuum or it maybe
takes place only in SWCNTs for the specific localization of
electrons (holes) at their surface and weak screening by the bound
electrons. That is why we consider here the simplest model of the
quasione-dimensional semiconductor with the cylindrical symmetry,
namely the linear crystal with two atoms in the unit cell. The
electrons (holes) in this crystal are simply localized at its
axis.

The aim of this work is only a qualitative analysis of the
mentioned effect. For study of electron structure of concerned 1D
crystal we apply here the zero-range potentials (ZRPs)
method~\cite{aghh},\cite{demk} (see section~2). The matter is that
results on the band structure and single-electron states, obtained
by this method for SWCNTs in~\cite{tish},\cite{adtish}, appeared
to be in good accordance with the experimental data and results of
ab initio calculations related to the band states. For certainty
we use the linear crystal parameters (the electron bare mass,
lattice parameters) taken from works on
nanotubes~\cite{tish},\cite{adtish}. In section~3 we obtain the
e-h bare interaction potential and that screened by the crystal
band electrons, and then the large-radius exciton spectrum for the
linear crystal in vacuum. All these data are used in section~4,
where we present results of calculations for the crystal with
different lattice periods (it also means different band
structures). As it turns out, the binding energy of even excitons
in the ground state well exceeds ($\sim2-5$ times) the energy gap
for the linear crystal in vacuum and the screening by the crystal
band electrons is negligible. Note, that this result was obtained
within the framework of exactly solvable ZRPs model with feasible
parameters. Therefore, the mentioned instability effect may take
place not only for the considered simplest case, but, most likely,
also for other quasione-dimensional isolated semiconductors in
vacuum.

\section{Single-electron band structure and eigenfunctions of band electrons}
\setcounter{equation}{0}

We have obtained the single-electron states in the linear crystal
using the zero-range potentials (ZRPs)
method~\cite{aghh},\cite{demk}. The main point of this method is
that the interaction of an electron with atoms or ions of a
lattice is described instead of some periodic potential
$V(\bi{r})$ by the sum of Fermi pseudo-potentials~\cite{aghh}
$(1/\alpha)\sum_l\delta(\rho_l)(\partial/\partial\rho_l)\rho_l$
($\rho_l=|\bi{r}-\bi{r}_l|$, $\bi{r}_l$ are the points of atoms
location, $\alpha$ is a certain fitting parameter) or equivalently
by the set of boundary conditions imposed on the single-electron
wave function $\psi$ at points $\bi{r}_l$:
\[
\lim_{\rho_l\rightarrow0}\left\{\frac{\rmd}{\rmd\rho_l}\left(\rho_l\psi\right)(\bi{r})+\alpha\left(\rho_l\psi\right)(\bi{r})\right\}=0.
\]
The electron wave functions satisfy at that the Schr\"{o}dinger
equation for a free particle for $\bi{r}\neq\bi{r}_l$. Therefore
we seek them for the linear crystal in the form:
\begin{equation}\label{2.1}
\psi\left(\rho^\mathrm{A}_n,\rho^\mathrm{B}_n\right)=\sum\limits_{n=-\infty}^\infty
A_n
\frac{\exp(-\kappa\rho^\mathrm{A}_n)}{\rho^\mathrm{A}_n}+\sum\limits_{n=-\infty}^\infty
B_n \frac{\exp(-\kappa\rho^\mathrm{B}_n)}{\rho^\mathrm{B}_n},
\end{equation}
where indices $\mathrm{A}$ and $\mathrm{B}$ denote two monatomic
sublattices of the diatomic lattice,
$\rho^\mathrm{A}_n=\left|\bi{r}-\bi{r}^\mathrm{A}_n\right|$ and
$\rho^\mathrm{B}_n=\left|\bi{r}-\bi{r}^\mathrm{B}_n\right|$, $n$
numbers all the sublattices points,
$\kappa=\sqrt{2m_\mathrm{b}|E|}/\hbar$, $E<0$ is the electron
energy and $m_\mathrm{b}$ is the bare mass. For certainty,
following~\cite{tish},\cite{adtish} we take from now on
$m_\mathrm{b}\simeq0.415m_\mathrm{e}$ and the ZRP parameter
$\alpha=\sqrt{2m_\mathrm{b}|E_\mathrm{ion}|}/\hbar$, where
$E_\mathrm{ion}$ is the ionization energy of an isolated carbon
atom. By~\cite{tish},\cite{adtish}, with these $\alpha$ and
$m_\mathrm{b}$ ZRPs method reproduces single-electron spectra of
such quasione-dimensional structures as SWCNTs within an accuracy
of existing experiments. One can take infinite limits for the
series in~(\ref{2.1}) even for the finite crystal, because terms
of these series decrease exponentially with increasing of $n$.

According to the ZRPs method the wave functions~(\ref{2.1}) should
satisfy the following boundary conditions at the all sublattices
points:
\begin{equation}\label{2.2}
\lim_{\rho^i_l\rightarrow0}\left\{\frac{\rmd}{\rmd\rho^i_l}\left(\rho^i_l\psi\right)(\bi{r})+\alpha\left(\rho^i_l\psi\right)(\bi{r})\right\}=0,
\end{equation}
here $i=\{\mathrm{A,B}\}$ according to each sublattice.

Further we suppose that the linear crystal lies along the
$z$-axis, thus $\bi{r}^\mathrm{A}_n=nd\bi{e}_\mathrm{z}$ and
$\bi{r}^\mathrm{B}_n=(nd+a)\bi{e}_\mathrm{z}$, where
$\bi{e}_\mathrm{z}$ is the $z$-axis unit vector, $a$ is the
distance between atoms in the unit cell of the crystal and $d>2a$
is the distance between the neighbour atoms in each sublattice.
Note, that $d=2a$ corresponds to the metallic monatomic crystal
and for the case $d<2a$ the smallest distance between atoms in the
crystal is $d-a<a$.

Substituting~(\ref{2.1}) to~(\ref{2.2}) and applying the Bloch
theorem ($A_n=A\exp(\rmi qdn)$, $B_n=B\exp(\rmi qdn)$, $q$ is the
electron quasi-momentum) we get two equations for amplitudes
$A,B$:
\begin{equation}\label{2.3}
\left\{ \begin{array}{c}
AQ_1+BQ_2=0,\\
AQ^*_2+BQ_1=0,\\
\end{array}\right.
\end{equation}
where
\begin{equation}\label{2.4}
Q_1(\kappa,q)=\alpha-\frac{1}{d}\ln\left(2[\cosh{\kappa
d}-\cos{qd}]\right),
\end{equation}
\begin{equation}\label{2.5}
Q_2(\kappa,q)=\sum\limits_{n=-\infty}^\infty\frac{\exp(-\kappa|nd+a|+\rmi
qnd)}{|nd+a|}.
\end{equation}
Setting $d=ja$:
\begin{equation}\label{2.6}
Q_2(\kappa,q)=\frac{1}{a}\int\limits_0^1\left(\frac{\exp[-\kappa
a]}{1-x^j\exp[d(\rmi q-\kappa)]}+\frac{x^{j-2}\exp[\kappa
a]}{\exp[d(\rmi q+\kappa)]-x^j}\right)\rmd x
\end{equation}
for each real $j>2$.

From~(\ref{2.3}) we get two equations, which define the band
structure of the crystal:
\begin{equation}\label{2.7}
Q_1(\kappa_1,q)-|Q_2(\kappa_1,q)|=0,
\end{equation}
\begin{equation}\label{2.8}
Q_1(\kappa_2,q)+|Q_2(\kappa_2,q)|=0.
\end{equation}
Equation~(\ref{2.7}) defines the conduction band and
equation~(\ref{2.8}) defines the valence band (see section~4,
figure~1). So the electron and hole effective masses can be simply
obtained from~(\ref{2.7}) and~(\ref{2.8}), respectively.

Further, using the Hilbert identity for Green's function of the 3D
Helmholtz equation, we obtain the normalized wave
functions~(\ref{2.1}):
\begin{equation}\label{2.9}
\begin{split}
\psi_{\kappa,q}(\bi{r})=&\frac{A(\kappa,q)}{\sqrt{L}}\left(\sum\limits_{n=-\infty}^\infty\frac{\exp(-\kappa|\bi{r}-nd\bi{e}_\mathrm{z}|+\rmi
qnd)}{|\bi{r}-nd\bi{e}_\mathrm{z}|}\right.\\
&\left.-\frac{Q_1}{Q_2}\sum\limits_{n=-\infty}^\infty\frac{\exp(-\kappa|\bi{r}-(nd+a)\bi{e}_\mathrm{z}|+\rmi
qnd)}{|\bi{r}-(nd+a)\bi{e}_\mathrm{z}|}\right),
\end{split}
\end{equation}
where $L$ is the crystal length and $A(\kappa,q)$ is the
normalization factor:
\[
A(\kappa,q)=\frac{1}{2}\left(\frac{\kappa
d}{\pi}\frac{\cosh{\kappa d}-\cos{qd}}{\sinh{\kappa d}-\Re
y}\right)^{1/2},
\]
and
\[
y=\frac{Q_1}{Q_2}\left(\exp[-\rmi qd]\sinh{\kappa
a}+\sinh{\kappa[d-a]}\right).
\]

\section{Exciton spectrum and eigenfunctions. Bare and screened e-h interaction}
\setcounter{equation}{0}

Using the same arguments as in the 3D case one can show (see, for
example~\cite{adsmyr}), that the wave equation for the Fourier
transform $\phi$ of envelope function in the wave packet from
products of the electron and hole Bloch functions, which
represents a two-particle state of large-radius rest exciton in a
(quasi)one-dimensional semiconductor with period $d$, is reduced
to the following 1D Schr\"{o}dinger equation:
\begin{equation}\label{3.1}
-\frac{\hbar^{2}}{2\mu}\phi''(z)+V(z)\phi(z)=\mathcal{E}\phi(z),\
\mathcal{E}=E_\mathrm{exc}-E_\mathrm{g}, \  -\infty<z<\infty ,
\end{equation}
where $\mu$ is the e-h reduced effective mass and $V(z)$ is the
e-h interaction potential:
\begin{equation}
\begin{split}
V(z)=-\int\limits_{\mathrm{E}_3^d}\int\limits_{\mathrm{E}_3^d}&\frac{e^2}{\left((x_1-x_2)^2+(y_1-y_2)^2+(z+z_1-z_2)^2\right)^{1/2}}\\
&\times|u_{\mathrm{c};\kappa,\pi/d}(\bi{r}_1)|^2|u_{\mathrm{v};\kappa,\pi/d}(\bi{r}_2)|^2\rmd\bi{r}_1\rmd\bi{r}_2,\\
&\mathrm{E}_3^d=\mathrm{E}_2\times(0<z<d).\nonumber
\end{split}
\end{equation}
Here $u_{\mathrm{c,v};\kappa,q}(\bi{r})$ are the Bloch amplitudes
of the Bloch wave functions
$\psi_{\mathrm{c,v};\kappa,q}(\bi{r})=\exp(\rmi
qz)u_{\mathrm{c,v};\kappa,q}(\bi{r})$ of the conduction and
valence band electrons of the linear crystal, respectively. Using
the actual localization of the Bloch amplitudes at the crystal
axis, after several Fourier transformations and simplifications we
adduce the e-h interaction potential to the following form:
\begin{equation}\label{3.2}
\begin{split}
V_{r_1,r_2}(z)=-\frac{4e^2r^2_1}{r_2d^2}\int\limits^\infty_0&\frac{J_1(k)J_1(kr_2/r_1)}{k^4}\left(\frac{k}{r_1}(|d-z|+|d+z|-2|z|)\right.\\
&\left.+\exp\left[-\frac{k}{r_1}|d-z|\right]+\exp\left[-\frac{k}{r_1}|d+z|\right]-2\exp\left[-\frac{k}{r_1}|z|\right]\right)\rmd
k,
\end{split}
\end{equation}
where $J$ is the Bessel function of the first kind and $r_1$
($r_2$) is the radius of the electron (hole) wave functions
transverse localization
\[
r_{1,2}=\left(2\int\limits_{\mathrm{E}_2}r^2_\mathrm{2D}\int\limits^L_0\left|u_{\mathrm{c,v};\kappa,q}(z,\bi{r}_\mathrm{2D})\right|^2\rmd
z \rmd\bi{r}_\mathrm{2D}\right)^{1/2},
\]
where $\bi{r}_\mathrm{2D}$ is the transverse component of the
radius-vector, $q=\pi/d$ and
$\kappa=\kappa_{1,2}\left(\pi/d\right)$ correspond to the
conduction and valence bands edges at the energy gap (according
to~(\ref{2.7}) and~(\ref{2.8}), respectively).
Equation~(\ref{3.1}) with the potential given by~(\ref{3.2})
defines the spectrum of large-radius exciton in the linear,
diatomic crystal if the screening effect by the crystal electrons
is ignored. Actually, the screening of the potential~(\ref{3.2})
by the band electrons is insignificant.

Indeed, following the Lindhard method (so-called RPA), to obtain
the e-h interaction potential $\varphi(\bi{r})$, screened by the
electrons of linear lattice, let us consider the Poisson equation:
\begin{equation}\label{3.3}
-\Delta\varphi(\bi{r})=4\pi\left(\rho^\mathrm{ext}(\bi{r})+\rho^\mathrm{ind}(\bi{r})\right),
\end{equation}
where $\bi{r}$ is the radius-vector, $\rho^\mathrm{ext}(\bi{r})$
is the density of extraneous charge and
$\rho^\mathrm{ind}(\bi{r})$ is the charge density induced by the
extraneous charge.

By~(\ref{3.3}) the screened e-h interaction potential may be
written as:
\begin{equation}\label{3.4}
\varphi(\bi{r})=4\pi\int\limits_\mathrm{E_3}\left(\rho^\mathrm{ext}(\bi{r}')+\rho^\mathrm{ind}(\bi{r}')\right)G(\bi{r},\bi{r}')\rmd\bi{r}',
\end{equation}
where $G(\bi{r},\bi{r}')=1/(4\pi|\bi{r}-\bi{r}'|)$ is Green's
function of the 3D Poisson equation.

Let $E^0(q)$ and $\psi^0_{\kappa,q}(\bi{r})=\exp(\rmi
qz)u^0_{\kappa,q}(\bi{r})$ be the band energies and corresponding
Bloch wave functions of the crystal electrons and $E(q)$,
$\psi_{\kappa,q}(\bi{r})$ be those in the presence of the
extraneous charge. Then
\begin{equation}\label{3.5}
\rho^\mathrm{ind}(\bi{r})=-e\sum\limits_{q}\left[f(E(q))|\psi_{\kappa,q}(\bi{r})|^2-f(E^0(q))|\psi^0_{\kappa,q}(\bi{r})|^2\right],
\end{equation}
where $f$ is the Fermi-Dirac function. Using the transverse
localization of the Bloch wave functions near the crystal axis, we
get in the linear in $\varphi$ approximation:
\begin{equation}\label{3.6}
\begin{split}
\rho^\mathrm{ind}(z',\bi{r}'_\mathrm{2D})=-e^2\sum\limits_{q,q'}\frac{1}{E_{\mathrm{g};q,q'}}\int\limits^{L}_{0}\int\limits_{\mathrm{E}_2}&u_{\mathrm{v};\kappa_2,q'}(z,\bi{r}_\mathrm{2D})u^*_{\mathrm{c};\kappa_1,q}(z,\bi{r}_\mathrm{2D})\rmd\bi{r}_\mathrm{2D}\varphi(z)\exp[\rmi
z(q'-q)]\rmd z\\
&\times
u^*_{\mathrm{v};\kappa_2,q'}(z',\bi{r}'_\mathrm{2D})u_{\mathrm{c};\kappa_1,q}(z',\bi{r}'_\mathrm{2D})\exp[\rmi
z'(q-q')],
\end{split}
\end{equation}
where $E_{\mathrm{g};q,q'}=E_{\mathrm{c}}(q)-E_{\mathrm{v}}(q')$.
Here and further $\varphi(z)$ is the e-h interaction potential
averaged in $\mathrm{E}_2$ over the region of the Bloch wave
functions transverse localization and over the lattice period $d$
along the crystal axis.

Due to the periodicity of the Bloch amplitudes $\rho^\mathrm{ind}$
may be written as:
\begin{equation}\label{3.7}
\begin{split}
\rho^\mathrm{ind}(\bi{r}')=-\frac{e^2N}{L}\sum\limits_{q,q'}&\frac{C(q,q';d)}{E_{\mathrm{g};q,q'}}\varphi(q-q')\\
&\times
u^*_{\mathrm{v};\kappa_2,q'}(\bi{r}')u_{\mathrm{c};\kappa_1,q}(\bi{r}')\exp[\rmi
z'(q-q')],
\end{split}
\end{equation}
where
\[
C(q,q';d)=\int\limits^d_0\int\limits_{\mathrm{E}_2}u^*_{\mathrm{c};\kappa_1,q}(z,\bi{r}_\mathrm{2D})u_{\mathrm{v};\kappa_2,q'}(z,\bi{r}_\mathrm{2D})\rmd\bi{r}_\mathrm{2D}\rmd
z
\]
and $N$ is the number of unit cells in the crystal.

Further, after several transformations we obtain from~(\ref{3.4})
and~(\ref{3.7}) the one-dimensional Fourier transform of the
potential $\varphi$:
\begin{equation}\label{3.8}
\begin{split}
&\varphi(k)=\frac{\varphi_0(k)}{\varepsilon(k)},\\
&\varepsilon(k)=1+\frac{e^2N^2}{2\pi^2}\int\limits^{\pi/d}_{-\pi/d}\frac{\left|C(q,q-k;d)\right|^2}{E_{\mathrm{g};q,q-k}}\rmd
q \widetilde{K}_0(k)\frac{2\sin(kd/2)}{kd},
\end{split}
\end{equation}
where $\varphi_0$ is the Fourier transform of the averaged
electrostatic potential induced by $\rho^\mathrm{ext}$ and
$\widetilde{K}_0(k)$ is the modified Bessel function of the second
kind averaged over $\bi{r}_\mathrm{2D}$ and $\bi{r}'_\mathrm{2D}$
in the region of the Bloch wave functions transverse localization
in $\mathrm{E}_2$, namely
\begin{equation}
\begin{split}
&\widetilde{K}_0(k)=\frac{1}{(\pi r_1r_2)^2}\int\limits_{\mathrm{E}^{r_1}_2}\int\limits_{\mathrm{E}^{r_2}_2}K_0(|k||\bi{r}_\mathrm{2D}-\bi{r}'_\mathrm{2D}|)\rmd\bi{r}_\mathrm{2D}\rmd\bi{r}'_\mathrm{2D},\\
&\mathrm{E}^{r_i}_2=(0\leq r_\mathrm{2D}\leq
r_i)\times(0\leq\beta\leq2\pi).\nonumber
\end{split}
\end{equation}

In the long-wave limit we get:
\begin{equation}\label{3.9}
\begin{split}
&\left|C(q,q-k;d)\right|^2_{k\rightarrow0}\approx\left|U(q;d)\right|^2k^2,\\
&U(q;d)=\int\limits^d_0\int\limits_{\mathrm{E}_2}u^*_{\mathrm{c};\kappa_1,q}(z,\bi{r}_\mathrm{2D})\frac{\partial}{\partial
q}u_{\mathrm{v};\kappa_2,q}(z,\bi{r}_\mathrm{2D})\rmd\bi{r}_\mathrm{2D}\rmd
z.
\end{split}
\end{equation}
Using of the Schr\"{o}dinger equation for the orthogonal Bloch
wave functions $\psi_{\kappa,q}(\bi{r})$ yields
\begin{equation}\label{3.10}
U(q;d)=\frac{\rmi\hbar^2}{E_{\mathrm{g};q,q}m_\mathrm{b}}\int\limits^d_0\int\limits_{\mathrm{E}_2}\psi^*_{\mathrm{c};\kappa_1,q}(z,\bi{r}_\mathrm{2D})\frac{\partial}{\partial
z}\psi_{\mathrm{v};\kappa_2,q}(z,\bi{r}_\mathrm{2D})\rmd\bi{r}_\mathrm{2D}\rmd
z.
\end{equation}

Hence, in the long-wave limit the screened quasione-dimensional
electrostatic potential induced by a charge $e_0$, distributed
with the density:
\begin{equation}
\begin{split}
&\rho^\mathrm{ext}(z,\bi{r}_\mathrm{2D})=\frac{e_0}{\pi
R^2d}(\Theta\left[z+d/2\right]-\Theta\left[z-d/2\right])(\Theta\left[r_\mathrm{2D}\right]-\Theta\left[r_\mathrm{2D}-R\right]),\ R>0,\\
&\Theta[x-x_0]=\left\{\begin{array}{c}0,\ x<x_0,\\
1,\ x>x_0,\nonumber\\
\end{array}\right.
\end{split}
\end{equation}
in accordance with~(\ref{3.8}) and~(\ref{3.10}), is given by the
expression
\begin{equation}\label{3.11}
\varphi(z)=\frac{8e_0r_1}{\pi
d^2}\int\limits^{\infty}_{0}\frac{(1/k^2)\sin^2(kd/2r_1)\widetilde{K}_0(k/r_1)\cos(kz/r_1)}{1+g_d
(kr_1/d)\sin(kd/2r_1)\widetilde{K}_0(k/r_1)}\ \rmd k
\end{equation}
with
\begin{equation}\label{3.12}
g_d=\left(\frac{e\hbar^2}{\pi r_1
m_\mathrm{b}}\right)^2\int\limits_{-\pi/d}^{\pi/d}\frac{1}{E^3_{\mathrm{g};q,q}}\left|\left\langle\psi_{\mathrm{c};\kappa_1,q}\left|\frac{\partial}{\partial
z}\right|\psi_{\mathrm{v};\kappa_2,q}\right\rangle\right|^2\rmd q.
\end{equation}

According to equation~(\ref{2.9}) the dimensionless screening
parameter $g_d$ may be also written as:
\begin{equation}\label{3.13}
\begin{split}
g_d=\left(\frac{16e}{\hbar
dr_1}\right)^2m_\mathrm{b}\int\limits_{0}^{\pi/d}&\frac{A^2_\mathrm{c}(\kappa_1,q)A^2_\mathrm{v}(\kappa_2,q)}{(\kappa^2_2(q)-\kappa^2_1(q))^5}\\
&\times\left|\left(1+\frac{Q_1(\kappa_1,q)Q_1(\kappa_2,q)}{Q^*_2(\kappa_1,q)Q_2(\kappa_2,q)}\right)\int\limits^{\kappa_2}_{\kappa_1}Q_1(\kappa,q)\rmd\kappa\right.\\
&-\left.\frac{Q_1(\kappa_1,q)}{Q^*_2(\kappa_1,q)}\int\limits^{\kappa_2}_{\kappa_1}Q^*_2(\kappa,q)\rmd\kappa-\frac{Q_1(\kappa_2,q)}{Q_2(\kappa_2,q)}\int\limits^{\kappa_2}_{\kappa_1}Q_2(\kappa,q)\rmd\kappa\right|^2\rmd
q.
\end{split}
\end{equation}
Note, that $\kappa_1$ and $\kappa_2$ are the implicit functions of
$q$ defined by~(\ref{2.7}) and~(\ref{2.8}), respectively.

It appears, that $g_d$ calculated according to~(\ref{3.13}) for
$d$ varying in the interval $[2.1a, 3a]$ are about $10^{-6}$.

\section{Discussion. Stabilization of single-electron states}
\setcounter{equation}{0}

Using equations~(\ref{2.7}) and~(\ref{2.8}) we obtained the band
structure (see figure~1) and the electrons and holes effective
masses for the linear crystal of dimers for different values of
the ratio $j=d/a$ of its period $d$ and the distance $a$ between
atoms in dimers. Besides, using wave equation~(\ref{3.1}) and
potentials~(\ref{3.2}) and~(\ref{3.11}) we found the large-radius
exciton energy spectrum in the crystal for the bare e-h
interaction and e-h interaction screened by the bound electrons of
the crystal. We present here results for the crystal with
$j\in[2.1,3]$. Contrary to the single-band metallic crystal with
$j=2$, the crystals with $j>2$ are semiconductors with band gaps
varying from zero to the difference between the electron levels in
an isolated dimer. Particularly, the crystals with $j=2.001$ and
realistic values of $a$ and $\alpha$ (as in nanotubes and some 1D
polymer chains) are narrow-gap semiconductors, in which excitons
may possess binding energies about $\sim10~\mathrm{meV}$, but the
crystals with $j=2.1$ are already wide-gap ($\sim1~\mathrm{eV}$)
semiconductors with strongly bound e-h pairs, and the crystals
with $j=3$ are almost flat band semiconductors, but their
electrons and holes at the energy gaps ($q=\pi/d$) still have the
finite effective masses (these electrons and holes form the
excitons in the crystals). For certainty, the distance $a$ we have
chosen equal to the graphite in-plane parameter
$0.142~\mathrm{nm}$. The ZRP interaction parameter
$\alpha=11.01~\mathrm{nm}^{-1}$ corresponds to the ionization
energy of an isolated carbon atom
($E_\mathrm{ion}=11.255~\mathrm{eV}$).

\begin{figure}[t]
\begin{center}
\includegraphics[scale=0.7]{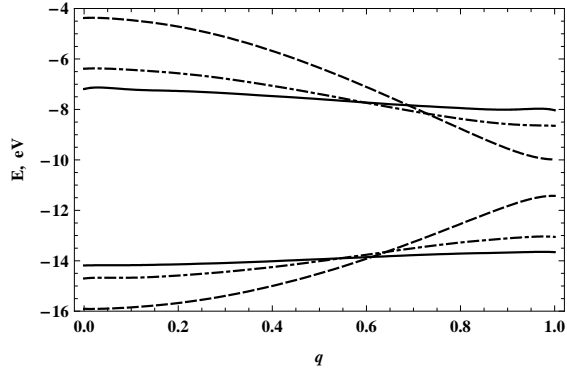}
\caption{The band structure of the linear crystal with parameters:
$j=2.1$ (dashed line), $j=2.5$ (dot-dashed line) and $j=3$ (solid
line); $q$ in units of $\pi/d$. The upper and lower bands
correspond to equation~(\ref{2.7}) and~(\ref{2.8}), respectively.}
\end{center}
\end{figure}
\begin{table}[t]
\caption{Band gaps $E_\mathrm{g}$ and reduced effective masses
$\mu$ according to~(\ref{2.7}),~(\ref{2.8}); radii of the
electrons and holes transverse localization $r_1$ and $r_2$,
respectively; screening parameter $g_d$ according to~(\ref{3.13})
and the exciton binding energies $\mathcal{E}$ of the even and odd
series for the linear, diatomic crystal according to
equation~(\ref{3.1}) with potential~(\ref{3.2}) for different
values of the ratio $j=d/a$.}
\begin{center}
\begin{tabular}{|l|l|l|l|l|l|l|l|l|}
\hline
$j$&$E_\mathrm{g}~\mathrm{(eV)}$&$\mu~(m_\mathrm{e})$&$r_1~\mathrm{(nm)}$&$r_2~\mathrm{(nm)}$&$g_d~(10^{-6})$&$\mathcal{E}_{0;\mathrm{even}}~\mathrm{(eV)}$&$\mathcal{E}_{1;\mathrm{odd}}~\mathrm{(eV)}$&$\mathcal{E}_{0;\mathrm{even}}/E_\mathrm{g}$\\
\hline
2.1&1.4422&0.041&0.070&0.0611&0.7235&-6.90&-0.5939&4.7845\\
\hline
2.3&3.3146&0.125&0.080&0.0569&2.4716&-8.9992&-2.0631&2.715\\
\hline
2.5&4.403&0.2199&0.088&0.0549&2.7036&-9.5352&-3.4812&2.1656\\
\hline
3&5.6281&0.5665&0.0994&0.0551&1.1206&-9.6588&-5.8421&1.7162\\
\hline
\end{tabular}
\end{center}
\end{table}

As one can see from table~1 the obtained from~(\ref{3.13})
dimensionless screening parameter $g_d\ll1$ for the all considered
values of $j$. So, it turns out that the screening of the e-h
interaction potential by the band electrons in the linear,
diatomic crystal may be ignored. This result could be expected
since the considered model of linear crystal is close to that of
the electron gas confined to a cylindrical well. In the latter
case, for which the separation of the angular variables takes
place, the states with different quantum numbers $m$ of the
angular momentum play the role of electron bands. Accordingly, the
matrix element
$\left|\left\langle\psi_\mathrm{c}\left|\partial/\partial
z\right|\psi_\mathrm{v}\right\rangle\right|$ from~(\ref{3.12}) for
the direct transitions between bands with different $m$ appears to
be identically equal to zero. This is why only the binding
energies of excitons with unscreened interaction potential are
listed in table~1.

To obtain estimates of the main linear crystal characteristics we
considered several limiting cases. In the case of $d\gg a$ (or
$j\gg1$) and $a=\mathrm{const}$ equations~(\ref{2.7})
and~(\ref{2.8}) can be reduced to
$\alpha-\kappa_{1,2}\mp\exp[-\kappa_{1,2}a]/a=0$, thus bands
become flat ($\kappa_{1,2}$ do not depend on $q$) and the band gap
tends to the finite value
$(\hbar^2/2m_\mathrm{b})(\kappa_2^2-\kappa_1^2)$ (for $a=0.142$~nm
it is about 6.3~eV), hence the reduced effective mass and exciton
binding energy tend to infinity, while the exciton radius
$r_\mathrm{exc}\sim\hbar^2/\mu e^2$ tends to zero. Therefore, the
large-radius exciton theory is actually appropriate
($r_\mathrm{exc}\gg a$) for excitons in the linear, diatomic
crystal only when its period $d$ runs the interval $(2a,2.4a)$
(e.g., $r_\mathrm{exc}$ is $\sim9a$ for $j=2.1$ and $\sim2a$ for
$j=2.4$). If $d=\mathrm{const}$, but $a\rightarrow0$, the
conduction band moves to the region of positive energies and at
some critical value of $a$ disappears, while the valence band
shifts correspondingly to the region of deep negative energies.

Table~1 shows that the ground-state exciton binding energies for
the linear, diatomic crystals with any value of the ratio $j$ are
greater than the corresponding energy gaps. Note, that according
to the same calculations, but with the bare mass
$m_\mathrm{b}=m_\mathrm{e}$, the ground-state exciton binding
energy for the linear crystal in vacuum appears significantly
greater than the energy gap. We should note also, that the
ground-state binding energies of excitons in the linear crystal
with different periods $d$ in vacuum, calculated using the 1D
analogue of the Ohno potential~\cite{ohnok} instead of
potential~(\ref{3.2}), remain greater than the corresponding
energy gaps. Particularly, for $d=2.3a$ calculations with the 1D
unscreened Ohno potential with the energy parameter $U$ taken
from~\cite{jiang} ($U=11.3~\mathrm{eV}$) and~\cite{cap}
($U=16~\mathrm{eV}$) give the ground-state exciton binding
energies $\mathcal{E}_{0;\mathrm{even}}=5.90~\mathrm{eV}$ and
$\mathcal{E}_{0;\mathrm{even}}=7.63~\mathrm{eV}$, respectively,
while $E_\mathrm{g}=3.31~\mathrm{eV}$ (see table~1). Thus, all
calculations made on the base of solvable zero-range potentials
model indicate the instability of the single-electron states in
the vicinity of the energy gap with respect to the formation of
excitons. This might be a shortage of this model, but it is worth
mentioning that results obtained on one-particle states in real 1D
systems, like SWCNTs~\cite{tish},~\cite{adtish}, within its
framework agree with existing experimental data in limits of
accuracy of the latter.

The stability of single-electron states of 1D semiconductors with
respect to the exciton formation in vacuum can be explained by
bringing multi-particle contributions into the picture. With the
advent of some number of excitons in the quasione-dimensional
crystal the additional screening appears, which is caused by a
rather great polarizability of excitons in the longitudinal
electric field. This collective contribution of born excitons into
the crystal permittivity returns the lowest exciton binding energy
$\mathcal{E}_{0;\mathrm{even}}$ into the energy gap and so blocks
further spontaneous transitions to the exciton states. To show
this let us consider the model of linear crystal immersed into the
gas of excitons with dielectric constant
$\varepsilon_\mathrm{exc}$ confined to the region of linear
crystal carriers transverse localization, namely: cylinder with
radius $r_1$ and axis coinciding with that of linear crystal (from
now on, for estimates, we assume that electron and hole have the
same transverse localization radius). In this case it is easy to
show that the e-h interaction potential is given by:
\begin{equation}\label{4.1}
\begin{split}
\varphi(z)=\frac{16e^2r_1}{\pi
d^2}\int\limits^{\infty}_{0}&\frac{\sin^2(kd/2r_1)\cos(kz/r_1)}{\varepsilon_\mathrm{exc}k^4}\\
&\times\left(1-\frac{2K_1(k)I_1(k)}{k(\varepsilon_\mathrm{exc}K_0(k)I_1(k)+K_1(k)I_0(k))}\right)\rmd
k,
\end{split}
\end{equation}
where $I_i$ and $K_i$ are the modified Bessel functions of the
order $i$ of the first and second kind, respectively. Further,
like in~\cite{jpcs}, we use the known elementary relation between
the permittivity of exciton gas and its polarizability $\alpha$ in
the direction of linear crystal
\[
\varepsilon_\mathrm{exc}=1+4\pi\alpha,\quad\alpha=2e^2n\sum\limits_k\frac{|\langle\Psi_0|\bi{r}|\Psi_k\rangle|^2}{\mathcal{E}_0-\mathcal{E}_k},
\]
where $n$ is the bulk concentration of excitons, $\Psi_0$ and
$\mathcal{E}_0$ are the exciton eigenfunction and binding energy,
which correspond to the ground state, and $\Psi_k$ and
$\mathcal{E}_k$ are those, which correspond to the all excited
states of exciton. Thus, the upper and lower limits for $\alpha$
are:
\[
\frac{2e^2n}{\mathcal{E}_0-\mathcal{E}_1}|\langle\Psi_0|\bi{r}|\Psi_1\rangle|^2\leq\alpha\leq\frac{2e^2n}{\mathcal{E}_0-\mathcal{E}_1}\sum\limits_k|\langle\Psi_0|\bi{r}|\Psi_k\rangle|^2=\frac{2e^2n}{\mathcal{E}_0-\mathcal{E}_1}|\langle\Psi_0|\bi{r}^2|\Psi_0\rangle|,
\]
where $\Psi_1$ and $\mathcal{E}_1$ correspond to the lowest
excited exciton state. Hence, one can obtain the upper and lower
limits for $n$:
\begin{equation}\label{4.2}
\begin{split}
\frac{\varepsilon_\mathrm{exc}-1}{4\pi}\frac{\mathcal{E}_{0;\mathrm{even}}-\mathcal{E}_{1;\mathrm{odd}}}{2e^2}\left|
\int\limits^\infty_{-\infty}z^2|\phi_0(z)|^2\rmd
z\right|^{-1}&\leq
n\\
&\leq\frac{\varepsilon_\mathrm{exc}-1}{4\pi}\frac{\mathcal{E}_{0;\mathrm{even}}-\mathcal{E}_{1;\mathrm{odd}}}{2e^2}\left|
\int\limits^{\infty}_{-\infty}z\phi_0(z)\phi_1(z)\rmd
z\right|^{-2},
\end{split}
\end{equation}
where each $\phi$ is the component of Fourier transform of the
corresponding exciton envelope function, it depends only on the
distance $z$ between the electron and hole. At that, $\phi_0$ is
the even solution of~(\ref{3.1}) with potential~(\ref{4.1}), which
corresponds to the exciton ground state and satisfies the boundary
condition $\phi'(0)=0$, and $\phi_1$ is the odd solution of the
same equation, which corresponds to the lowest excited exciton
state and satisfies the boundary condition $\phi(0)=0$.

Varying $\varepsilon_\mathrm{exc}$ in~(\ref{4.1}) substituted into
the wave equation~(\ref{3.1}) one can match
$\mathcal{E}_{0;\mathrm{even}}$ to the forbidden band width.
Further, $\mathcal{E}_{1;\mathrm{odd}}$ can be obtained from the
same equation with the fixed $\varepsilon_\mathrm{exc}$ and with
the corresponding boundary condition. All these magnitudes allow
to calculate from~(\ref{4.2}) the rough upper and lower limits for
the critical concentration of born excitons $n_\mathrm{c}$, which
is sufficient to return $\mathcal{E}_{0;\mathrm{even}}$ into the
energy gap. Further, knowing $n_\mathrm{c}$ we can calculate the
shift of the forbidden band edges, which move apart due to the
transformation of some single-electron states into excitons. This
results in the enhancement of energy gap
\begin{equation}\label{4.3}
\delta
E_\mathrm{g}\simeq\frac{(\pi\hbar\widetilde{n}_\mathrm{c})^2}{2\mu}
\end{equation}
like in~\cite{ohno} and~\cite{kiow}. Here
$\widetilde{n}_\mathrm{c}=n_\mathrm{c}\pi r_1^2$ is the linear
critical concentration of excitons, and $r_1$ is the radius of
electron wave functions transverse localization at the linear
crystal axis.

In accordance with~(\ref{4.2}) the described model yields
$\widetilde{n}_\mathrm{c}\sim180~\mathrm{\mu m}^{-1}$ ($\sim3\%$
of the atoms number in the crystal) and $\sim400~\mathrm{\mu
m}^{-1}$ ($\sim7\%$) for the linear crystal with $j=2.1$ and
$j=2.3$, respectively, while by~(\ref{4.3}) the corresponding
$\delta E_\mathrm{g}$ are $\sim300~\mathrm{meV}$ and
$\sim500~\mathrm{meV}$ in the same order. Here, however, we should
mention that for SWCNTs both the measured
in~\cite{ohno},\cite{kiow} and estimated in the same
manner~\cite{jpcs} values of $\delta E_\mathrm{g}/E_\mathrm{g}$
appeared to be two-four times less.

Note, finally, that the instability of the single-electron states
weakens or disappears for linear crystals immersed into dielectric
media. As it was shown in~\cite{ppv} by the example of the
poly-para-phenylenevinylene 1D chain or
in~\cite{jiang},\cite{cap},\cite{pereb} by the example of SWCNTs
the environmental effect may substantially decrease the excitons
binding energies. Indeed, for the linear crystal in media even
with permittivity about $\varepsilon\sim2-3$ (e.g., like
in~\cite{jiang} or~\cite{cap}) the ground-state exciton binding
energy becomes smaller than the energy gap.

\section*{Acknowledgments}
This work was supported by the Ministry of Education and Science
of Ukraine, Grant \#0106U001673.

\end{document}